\documentclass[11pt,letterpaper]{article}
\usepackage{amsmath}
\usepackage{amssymb}
\usepackage[dvips]{graphicx}
\begin{document}
\title{Potential-density pairs for bent bars}
\author{D. Vogt\thanks{e-mail: dvogt@ime.unicamp.br} 
\and
P. S. Letelier\thanks{e-mail: letelier@ime.unicamp.br}\\
Departamento de Matem\'{a}tica Aplicada-IMECC, Universidade \\
Estadual de Campinas 13083-970 Campinas, S\~ao Paulo, Brazil}
\maketitle
\begin{abstract}
A method is presented to bend a thin massive line when the curvature 
is small. The procedure is applied to a homogeneous thin bar with two types of curvatures. 
One of them mimics a galactic bar with two spiral arms at its tips. It is showed that if the 
bending function is a linear combination of Legendre polynomials, then the bent potential 
is an exact solution of the Laplace equation. A transformation is applied on the thin bent 
bars to generate three-dimensional potential-density pairs without singularities. The potentials
of the thin bent bars are also used to generate non-axisymmetric planar distributions of matter.

\end{abstract}

\section{Introduction}
Bars are a common self-gravitating structure present in disc galaxies. About 50 per cent of such galaxies
are stongly or weakly barred, including our Milky Way \cite{sw93,bm98}; see also the classification of 
galaxies by \cite{dev63} and the fraction of barred galaxies discussed by \cite{kn99,esk00,ksp00}. Galactic bars are 
triaxial systems, and constructing analytical triaxial potential-density pairs is a difficult task. 
The only exact, self-consistent models of bars were constructed by Freeman \cite{fr66}, but they 
have some unrealistic features for barred systems. As alternatives, galactic bars have been approximately modelled 
as homogeneous ellipsoids \cite{d65,m75} or inhomogeneous prolate spheroids \cite{vf72,abmp83,pp83,pf84}. 
In these works, the inhomogeneous bar has been represented by a Ferrers ellipsoid \cite{fe77}, which has a finite length and 
represents many features of galactic bars rather well. Long \& Murali \cite{lm92} found simple analytical potential density-pairs for 
prolate and triaxial bars that can all be expressed in terms of elementary functions. One of their model of bar was used 
 in hydrodynamic simulations \cite{llak99,al00,al04,at05,taj09} to understand the response of a gaseous disc to the 
imposition of non-axisymmetric bar potentials. 

Until recently, an unclear issue was the
connection between bars and grand-design spirals. Many barred galaxies have spiral arms that 
appear to emerge from the tips of the bar (see, e.g., NGC 1300; \cite{bt08}, p.\ 525). 
However, \cite{ss88} present evidences that the pattern speed of 
the spirals are much lower than the pattern speed of the bar, so the spiral cannot be driven directly 
by the bar. On the other hand, there are other observational evidences that bars and spiral arms are 
correlated \cite{ee89,bl01,bl04,bu09}.  Salo et al. \cite{slbk10} recently investigated the relation between bar forcing and spiral
density amplitudes for over 100 barred galaxies, and found that there exists a significant statistical correlation. 
Furthermore, hydrodynamics simulations of the response of a gaseous disc to the imposition of a 
non-axisymmetric bar \cite{ath92,w94,eg97,fwh98,al00,es00,pa00,ma02,ma03,at05} have shown that the 
symmetric two-armed spirals in barred galaxies are driven by 
the gravitational torques of the bar. 
  
The system bar+spiral arms may be viewed as a bar with bended ends, although 
we are not aware that such an interpretation has been proposed so far.  
It would be interesting to have simple, analytical models for such a gravitating system.
In this work we propose a method to obtain potential-density pairs for thin and for `softened'
bent bars. In Section \ref{sec_lines}, we present a procedure to bend
a thin massive line and to calculate its potential. The idea is to
consider a slight curvature and expand the potential with respect to a
small parameter. In Section \ref{sec_bars} this formalism will be particularized 
to a thin bar with constant linear density. It will be shown that if the 
`bending function' can be written in terms of Legendre polynomials, 
then the potential of the bent bar will be an \emph{exact} solution of the
Laplace equation. Two examples of deformed bars will be discussed. By using a suitable 
transformation, the thin bent bars are then `softened'  to generate three-dimensional 
potential-density pairs without singularities. In 
Section \ref{sec_nax}, we present non-axisymmetric potential density-pairs
that represent planar distributions of matter constructed from
the two potentials of bent bars discussed in Section \ref{sec_bars}. 
These planar potential-density pairs are found by using a method first proposed by  
Kuzmin \cite{k56}. The discussion of the results is left to Section \ref{sec_dis}.

\section{Bent massive lines} \label{sec_lines}
In this section we present a procedure to bend a thin massive line.
The curvature is supposed to be small, and the potential
of the bent system will be obtained from an expansion of a small parameter.
The undeformed massive line will be described by the parametric equations 
$x^{\prime}=x_0(t)$, $y^{\prime}=y_0(t)$, $z^{\prime}=z_0(t)$, with $t \in [t_1,t_2]$.
The gravitational potential $\Phi(\mathbf{r})$ of this line can be expressed as
\begin{equation} \label{eq_pot_s}
\Phi=-G \int_{s_1}^{s_2} \frac{\lambda(s)\mathrm{d}s}{\lvert
\mathbf{r}-\mathbf{r^{\prime}} \rvert} \mbox{,}
\end{equation}
where $G$ is the gravitational constant, $s$ is the arc length
and $\lambda(s)$ is the linear density. In terms of the parameter 
$t$, equation (\ref{eq_pot_s}) is rewritten as
\begin{equation} \label{eq_pot_t}
\Phi=-G \int_{t_1}^{t_2} \frac{\lambda(t)\lvert \dot{\mathbf{r^{\prime}}} 
\rvert \mathrm{d}t}{\lvert \mathbf{r}-\mathbf{r^{\prime}} \rvert} \mbox{,}
\end{equation} 
where the dot represents derivative with respect to $t$.  

We shall represent a slight curvature of the line as
\begin{equation} \label{eq_def}
\mathbf{r^{\prime}}=\mathbf{r}_0(t)+\varepsilon \mathbf{r}_1(t) \mbox{,}
\end{equation}
where $\mathbf{r}_0(t)$ is the position of the undeformed line,
$\varepsilon$ is a small dimensionless parameter 
and the particular form of the curvature will be determined by
the function $\mathbf{r}_1(t)$. We have, up to first order in $\varepsilon$,
\begin{gather}
\lvert \dot{\mathbf{r^{\prime}}} \rvert = \lvert \dot{\mathbf{r}}_0 \rvert 
+\varepsilon \frac{\dot{\mathbf{r}}_0 \cdot \dot{\mathbf{r}}_1}
{\lvert \dot{\mathbf{r}}_0 \rvert} \mbox{,}\\
\frac{1}{\lvert \mathbf{r}-\mathbf{r^{\prime}} \rvert}= 
\frac{1}{\lvert \mathbf{r}-\mathbf{r}_0 \rvert} +\varepsilon 
\frac{\mathbf{r}_1 \cdot \left( \mathbf{r}-\mathbf{r}_0 \right)}
{\lvert \mathbf{r}-\mathbf{r}_0 \rvert^3} \mbox{.}
\end{gather}
Thus, up to first order in $\varepsilon$, the potential (\ref{eq_pot_t}) can be written as 
\begin{equation} \label{eq_pert_l}
\Phi=-G \int_{t_1}^{t_2} \frac{\lambda(t)\lvert \dot{\mathbf{r}}_0
\rvert \mathrm{d}t}{\lvert \mathbf{r}-\mathbf{r}_0 \rvert} 
-\varepsilon G \int_{t_1}^{t_2} \frac{\lambda(t) \dot{\mathbf{r}}_0 \cdot \dot{\mathbf{r}}_1 
\mathrm{d}t}{\lvert \dot{\mathbf{r}}_0 \rvert 
\lvert \mathbf{r}-\mathbf{r}_0 \rvert} - \varepsilon G 
\int_{t_1}^{t_2} \frac{\lambda(t)\lvert \dot{\mathbf{r}}_0
\rvert \mathbf{r}_1 \cdot \left( \mathbf{r}-\mathbf{r}_0 \right) \mathrm{d}t}
{\lvert \mathbf{r}-\mathbf{r}_0 \rvert^3} \mbox{.}
\end{equation}

\section{Bent bars} \label{sec_bars}
Now we consider the particular case of a bent bar with constant 
linear density. The undeformed bar located symmetrically on the $z$-axis with length $2a$
and linear density $\lambda_0$ will be parameterized by
$\mathbf{r}_0=(0,0,t)$, with $t \in [-a,a]$. For simplicity, the `bending function'
$\mathbf{r}_1(t)$ will be chosen as $\mathbf{r}_1=\bigl(0,f(t),0\bigr)$. With 
these assumptions, the potential (\ref{eq_pert_l}) reduces to 
\begin{equation} \label{eq_phi_d}
\Phi=\Phi_0+\varepsilon\Phi_1 \mbox{,} 
\end{equation}
where
\begin{gather} 
\Phi_0=-G\lambda_0 \int_{-a}^{a} \frac{\mathrm{d}t}{\sqrt{x^2+y^2+\left( z-t \right)^2}} \mbox{,} \notag \\ 
\Phi_1=-G\lambda_0 \int_{-a}^{a} \frac{yf(t) \mathrm{d}t}
{\left[ x^2+y^2+\left( z-t \right)^2 \right]^{3/2}} \mbox{.} \label{eq_phi01} 
\end{gather}

At this point it is convenient to relate the potentials (\ref{eq_phi01}) to 
an identity found by Letelier \cite{l99}:
\begin{equation} \label{eq_ide}
Q_n(u)P_n(v)=\frac{1}{2} \int_{-a}^{a} \frac{P_n(t/a)\mathrm{d}t}{\sqrt{R^2+
\left( z-t \right)^2}} \mbox{,}
\end{equation} 
where $P_n$ and $Q_n$ are, respectively, the Legendre polynomials and the Legendre 
functions of the second kind, and $(u,v)$ are the spheroidal coordinates related to
the cylindrical coordinates $(R,z)$ through 
\begin{gather}
u =(R_1+R_2)/(2a) \text{,} \qquad  v = (R_1-R_2)/(2a) \mbox{,} \\
R_1 =\sqrt{R^2+\left(z+a \right)^2} \text{,} \qquad  R_2 = \sqrt{R^2+\left(z-a \right)^2} \mbox{,}
\end{gather}
with $u \geq 1$ and $-1 \leq v \leq 1$. The physical interpretation of (\ref{eq_ide}) 
is in terms of a bar with linear density proportional to a Legendre polynomial, whose 
potential corresponds to a multipole term that arises in a multipolar solution of the
Einstein equations. The potential $\Phi_0$ in (\ref{eq_phi01}) can thus be identified with
\begin{equation} \label{eq_phi_0}
\Phi_0=-2G\lambda_0Q_0(u)P_0(v)=-2G\lambda_0Q_0(u) \mbox{.}
\end{equation}
Furthermore, we note that $\Phi_1$ can be rewritten as
\begin{equation} \label{eq_deriv}
\Phi_1=-G\lambda_0 \int_{-a}^{a} \frac{yf(t) \mathrm{d}t}
{\left[ x^2+y^2+\left( z-t \right)^2 \right]^{3/2}} = 
G\lambda_0 \frac{\partial}{\partial y} \int_{-a}^{a} \frac{f(t) \mathrm{d}t}
{\sqrt{x^2+y^2+\left( z-t \right)^2}} \mbox{.}
\end{equation}
If the function $f(t)$ can be expressed as a linear combination of Legendre
polynomials, then the last integral in (\ref{eq_deriv}) may be directly integrated via the
identity (\ref{eq_ide}). This also shows that the potential $\Phi_1$ will be then a solution
of the Laplace equation, thus the potential (\ref{eq_phi_d}) of the bent bar will also be an
exact solution of the Laplace equation.

\begin{figure}
\centering
\includegraphics[scale=0.5]{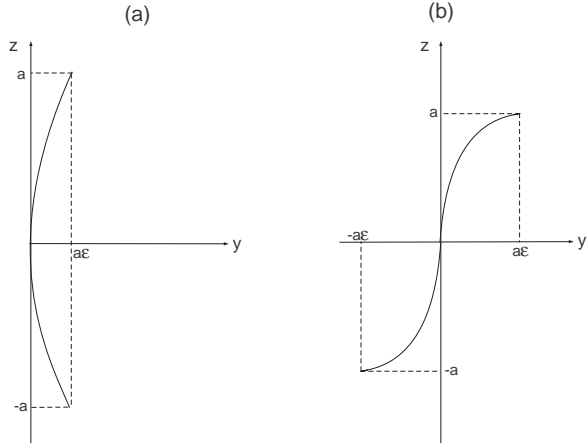}
\caption{The shape of the bent bar for (a) the function $f(t)=t^2/a$, and 
(b) the function $f(t)=t^3/a^2$.} \label{fig1}
\end{figure}

As examples, we will calculate the potential-density pairs for bars 
with two choices for the `bending function' $\mathbf{r}_1=\bigl(0,f(t),0\bigr)$. 
By (\ref{eq_def}), the parametric equations
of the deformed bar are $x^{\prime}=0$, $y^{\prime}=\varepsilon f(t)$ and $z^{\prime}=t$, $t \in [-a,a]$,
which may be rewritten as $y^{\prime}=\varepsilon f(z^{\prime})$, $-a \leq z^{\prime} \leq a$. We shall take as first example
$f(t)=t^2/a$ and then $f(t)=t^3/a^2$. The shape of the bent bar in each case is
schematically depicted in Figs \ref{fig1}(a) and (b), respectively. The first shape does not represent 
necessarily a galactic bar, but we chose this bending function for its simplicity. The second shape mimics 
a bar with two spiral arms.

In terms of Legendre polynomials, we have 
\begin{align}
\frac{t^2}{a} &=\frac{a}{3}P_0\left(\frac{t}{a}\right)+\frac{2a}{3}P_2\left(\frac{t}{a}\right) \mbox{,} \\
\frac{t^3}{a^2} &=\frac{3a}{5}P_1\left(\frac{t}{a}\right)+\frac{2a}{5}P_3\left(\frac{t}{a}\right) \mbox{.}
\end{align}
It is then straightforward to calculate the potential (\ref{eq_deriv}) using
the identity (\ref{eq_ide}), and we state the final result
\begin{gather}
\Phi_{(a)}=G\lambda_0 \ln \left( \frac{z-a+R_2}{z+a+R_1} \right) +\varepsilon
\frac{G\lambda_0y}{a} \left[ \ln \left( \frac{z-a+R_2}{z+a+R_1} \right) \right. \notag \\
\left. +\frac{a\left( x^2+y^2-z^2\right)\left( R_1+R_2 \right)+ 
z\left( x^2+y^2+z^2\right)\left( R_1-R_2 \right)}{\left( x^2+y^2\right)R_1R_2} \right] \mbox{,}
\label{eq_phi_a1} \\
\Phi_{(b)}= G\lambda_0 \ln \left( \frac{z-a+R_2}{z+a+R_1} \right) 
+\varepsilon \frac{G\lambda_0y}{a^2\left( x^2+y^2\right)R_1R_2} \notag \\
\times \left\{ 3z\left( x^2+y^2\right)R_1R_2 
\ln \left( \frac{z-a+R_2}{z+a+R_1} \right)+az\left( 5x^2+5y^2-z^2\right)\left( R_1+R_2 \right) \right. \notag \\
\left. -\left( R_1-R_2 \right)\left[ a^2\left( x^2+y^2 \right)+\left( x^2+y^2+z^2\right)
\left(  2x^2+2y^2-z^2 \right)\right] \right\} \mbox{,} \label{eq_phi_b1}
\end{gather}
where $R_1 =\sqrt{x^2+y^2+\left(z+a \right)^2}$, $R_2 =\sqrt{x^2+y^2+\left(z-a \right)^2}$  
and the subscripts $(a)$ and $(b)$ refer to the potentials calculated with the 
functions $f(t)=t^2/a$ and $f(t)=t^3/a^2$,
respectively. The potential (\ref{eq_phi_a1}) remains invariant under the transformations
$x \rightarrow -x$ or $z \rightarrow -z$, whereas the potential (\ref{eq_phi_b1}) remains
invariant under the transformations $x \rightarrow -x$ or $y \rightarrow -y,z \rightarrow -z$.

\subsection{Softened bent bars}
The potentials (\ref{eq_phi_a1})--(\ref{eq_phi_b1}) become singular along the bent
thin bars. In order to get physically more realistic potentials it is convenient to `soften' 
them. A simple way to achieve this is by means of a Plummer-like transformation \cite{lm92}. 
In our examples we apply a transformation $x^2+y^2 \rightarrow x^2+y^2+b^2$, where $b>0$ 
is a `softening' parameter. The corresponding three-dimensional mass density distribution 
is obtained from the Poisson equation in Cartesian coordinates, $\nabla^2 \Phi(x,y,z)=4\pi G \rho$.
From the potential (\ref{eq_phi_a1}), after the transformation, we obtain the pair
\begin{gather} 
\Phi_{(a)}=G\lambda_0 \ln \left( \frac{z-a+\mathcal{R}_2}{z+a+\mathcal{R}_1} \right) +\varepsilon
\frac{G\lambda_0y}{a} \left[ \ln \left( \frac{z-a+\mathcal{R}_2}{z+a+\mathcal{R}_1} \right) \right. \notag \\
\left. +\frac{a\left( x^2+y^2+b^2-z^2\right)\left( \mathcal{R}_1+\mathcal{R}_2 \right)+
z\left( x^2+y^2+b^2+z^2\right)\left( \mathcal{R}_1-\mathcal{R}_2 \right)}
{\left( x^2+y^2+b^2\right)\mathcal{R}_1\mathcal{R}_2} \right] \mbox{,} \label{eq_phi_a2} \\
\rho_{(a)}=\rho_0+\varepsilon\rho_1 \mbox{,} \label{eq_rho_a}
\end{gather}
where
\begin{gather}
\rho_0=\frac{\lambda_0b^2}{4\pi \left( x^2+y^2+b^2\right)^2\mathcal{R}_1^3\mathcal{R}_2^3} 
\left\{ \mathcal{R}_2^3\left( z+a \right) \left[ 3\left( x^2+y^2+b^2\right)
+2\left( z+a \right)^2\right] \right. \notag \\
\left. -\mathcal{R}_1^3\left( z-a \right) 
\left[ 3\left( x^2+y^2+b^2\right)+2\left( z-a \right)^2\right] \right\} \label{eq_rho_0} \mbox{,} \\
\rho_1=\frac{\lambda_0yb^2}{4\pi a\left( x^2+y^2+b^2\right)^3\mathcal{R}_1^5\mathcal{R}_2^5}
\left\{ 6z\left( x^2+y^2+b^2\right)^3 \left( \mathcal{R}_2^5-\mathcal{R}_1^5 \right) \right. \notag \\
\left. - \mathcal{R}_1^5\left( z-a \right) \left[ 8z^2\left( z-a \right)^4 +2\left( z-a \right)^2 
\left( x^2+y^2+b^2\right)\left( 11z^2-2az+a^2\right) \right. \right. \notag \\
\left. \left. +5\left( x^2+y^2+b^2\right)^2 
\left( 4z^2-2az+a^2\right) \right] +\mathcal{R}_2^5\left( z+a \right)
\left[ 8z^2\left( z+a \right)^4 \right. \right. \notag \\ 
\left. \left. +2\left( z+a \right)^2
\left( x^2+y^2+b^2\right)\left( 11z^2+2az+a^2\right)  \right. \right. \notag \\ 
\left. \left. +5\left( x^2+y^2+b^2\right)^2 \left( 4z^2+2az+a^2\right) \right] \right\} \mbox{,}
\end{gather}
where $\mathcal{R}_1=\sqrt{x^2+y^2+b^2+\left(z+a \right)^2}$, and 
$\mathcal{R}_2=\sqrt{x^2+y^2+b^2+\left(z-a \right)^2}$. Note that the first term 
of the potential (\ref{eq_phi_a2}) is the same as the potential of the prolate 
 bar of \cite{lm92} with the replacements $z \rightarrow x$ and $x^2+y^2 
\rightarrow y^2+z^2=R^2$.  

\begin{figure}
\centering
\includegraphics[scale=0.75]{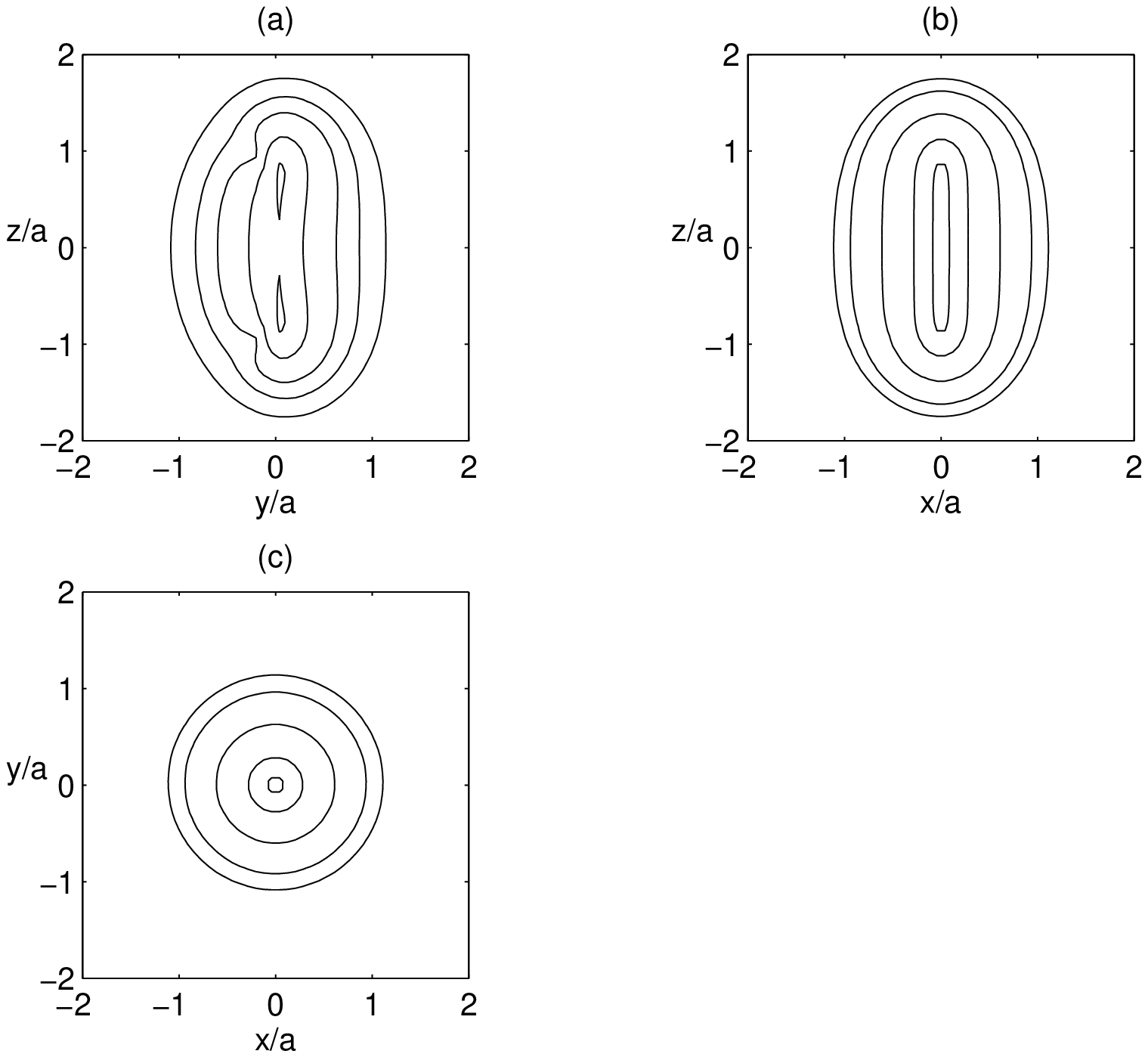}
\caption{Isodensity contours of the mass density $\bar{\rho}_{(a)}=\rho_{(a)}/(\lambda_0/a^2)$,
equation (\ref{eq_rho_a}), in the three orthogonal coordinate planes. Parameters: 
$b/a=0.25$, $\varepsilon=0.15$.} \label{fig2}
\end{figure}

\begin{figure}
\centering
\includegraphics[scale=0.75]{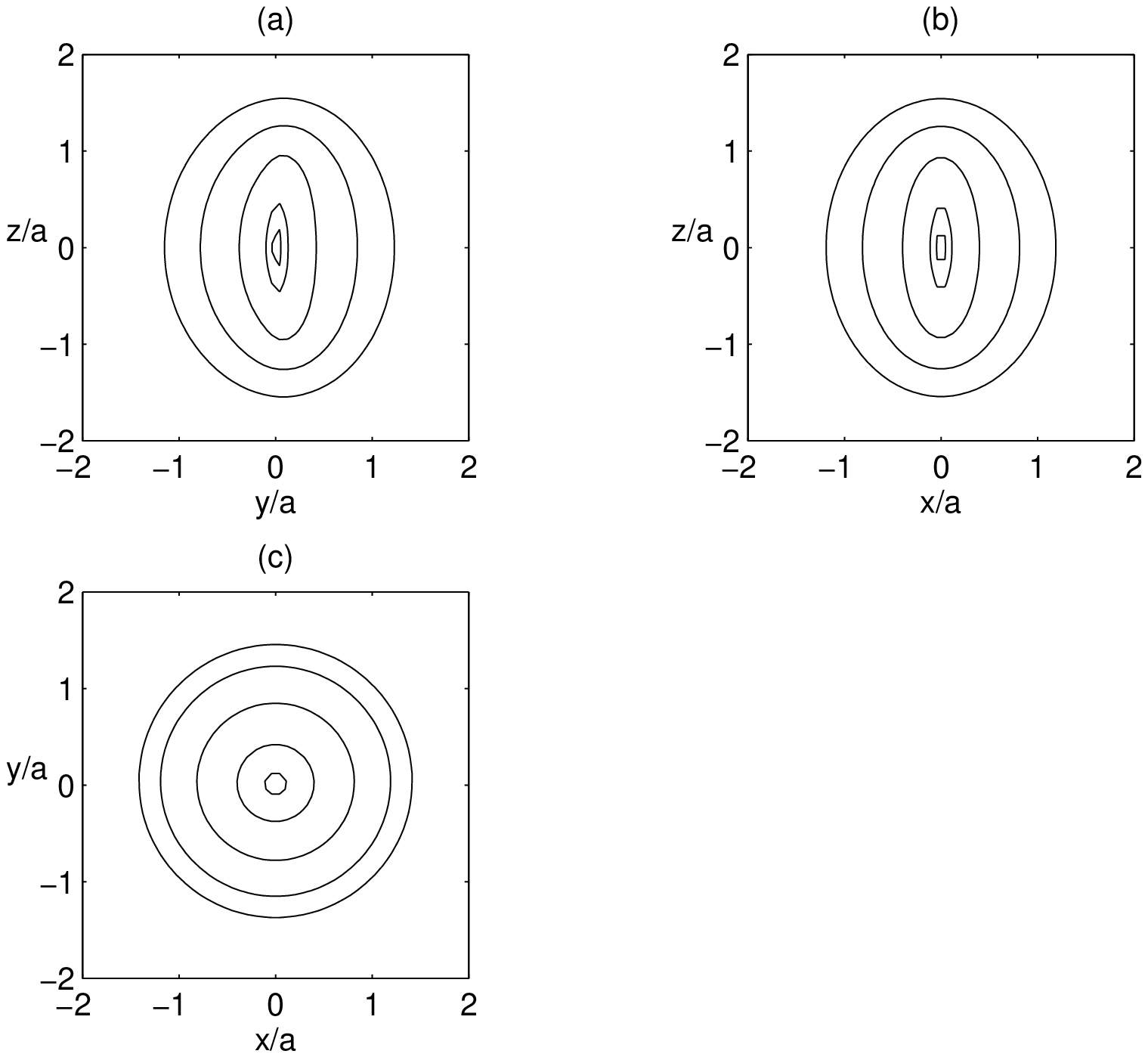}
\caption{Isopotential contours of the potential $\bar{\Phi}_{(a)}=\Phi_{(a)}/(G\lambda_0)$,
equation (\ref{eq_phi_a2}), in the three orthogonal coordinate planes. Parameters:
$b/a=0.25$, $\varepsilon=0.15$.} \label{fig3}
\end{figure}

\begin{figure}
\centering
\includegraphics[scale=0.75]{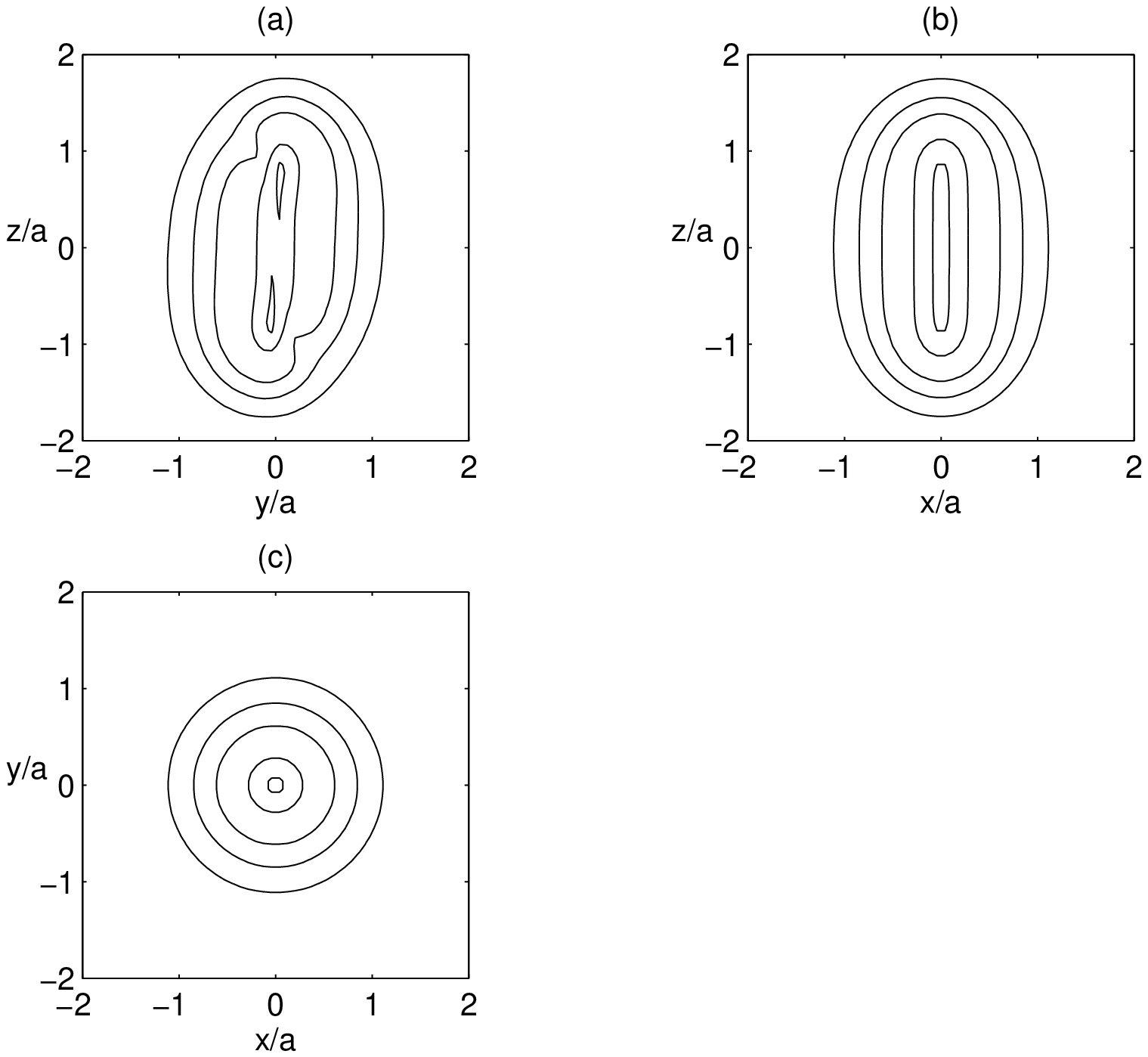}
\caption{Isodensity contours of the mass density $\bar{\rho}_{(b)}=\rho_{(b)}/(\lambda_0/a^2)$,
equation (\ref{eq_rho_b}), in the three orthogonal coordinate planes. Parameters:
$b/a=0.25$, $\varepsilon=0.15$.} \label{fig4}
\end{figure}

\begin{figure}
\centering
\includegraphics[scale=0.75]{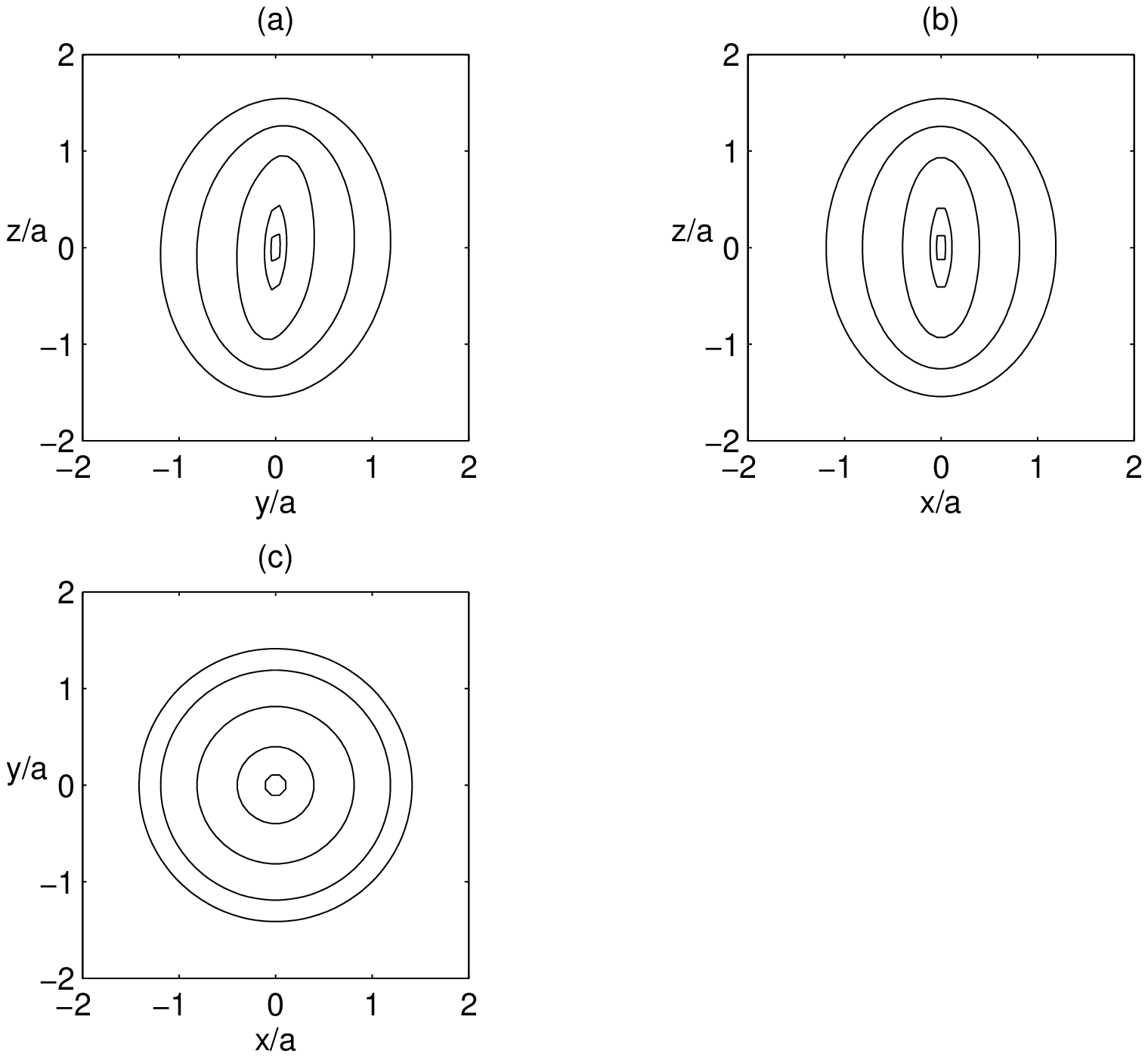}
\caption{Isopotential contours of the potential $\bar{\Phi}_{(b)}=\Phi_{(b)}/(G\lambda_0)$,
equation (\ref{eq_phi_b2}), in the three orthogonal coordinate planes. Parameters:
$b/a=0.25$, $\varepsilon=0.15$.} \label{fig5}
\end{figure}

Figs \ref{fig2}(a)--(c) display some curves of constant density of the dimensionless 
mass density $\bar{\rho}_{(a)}=\rho_{(a)}/(\lambda_0/a^2)$, equation (\ref{eq_rho_a}), 
in the three orthogonal planes, with parameters $b/a=0.25$ and $\varepsilon=0.15$. The bending of
the bar is apparent in the $y-z$ plane in Fig.\ \ref{fig2}(a). The density contours 
near the origin are bent in the same manner as the thin line of Fig.\ \ref{fig1}(a).
There are two points of maximum that are
shifted to the right of the $z$-axis, and two points of minimum appear on the opposite
side. At these points the mass density may become negative as $\varepsilon$ is
increased; we found that for $b/a=0.25$ the density is non-negative for
$\varepsilon \lessapprox 0.16$. As the parameter $b/a$ is increased for a fixed value of $\varepsilon$, 
the density contours become less elongated, the two points of minimum may disappear,
and the two points of maximum move towards the origin.
The contours in the $x-z$ plane preserve axial symmetry.
Some level curves of the potential $\bar{\Phi}_{(a)}=\Phi_{(a)}/(G\lambda_0)$, equation (\ref{eq_phi_a2}),
with parameters $b/a=0.25$ and $\varepsilon=0.15$, are displayed in Figs \ref{fig3}(a)--(c). 
In the $y-z$ plane the potential has only one minimum on $z=0$ and $y=y_m$, which 
is the zero of
\begin{multline} 
\frac{2ay_m}{\left( y_m^2+b^2\right)\sqrt{y_m^2+b^2+a^2}}+\frac{\varepsilon}{a} \ln 
\left( \frac{\sqrt{y_m^2+b^2+a^2}-a}{\sqrt{y_m^2+b^2+a^2}+a}\right) \\
+\frac{2\varepsilon \left[ \left( y_m^2+b^2 \right)^2+a^2\left( 2y_m^2+b^2\right)\right]}
{\left( y_m^2+b^2\right)\left( y_m^2+b^2+a^2\right)^{3/2}}=0 \mbox{.} \label{eq_y_m}
\end{multline}
For the above-mentioned values of parameters the minimum occurs at $y_m/a \approx 0.011$.  
The components of the force $\mathbf{F}_{(a)}=-\nabla \Phi_{(a)}$, 
corresponding to the potential (\ref{eq_phi_a2}),  are listed in Appendix \ref{ap_A}. 

Applying the same transformation on the potential (\ref{eq_phi_b1}) results in the 
potential-density pair: 
\begin{gather} 
\Phi_{(b)}= G\lambda_0 \ln \left( \frac{z-a+\mathcal{R}_2}{z+a+\mathcal{R}_1} \right) +\varepsilon
\frac{G\lambda_0y}{a^2\left( x^2+y^2+b^2\right)\mathcal{R}_1\mathcal{R}_2} \notag \\
\times \left\{ 3z\left( x^2+y^2+b^2\right)\mathcal{R}_1\mathcal{R}_2
\ln \left( \frac{z-a+\mathcal{R}_2}{z+a+\mathcal{R}_1} \right) \right. \notag \\
\left. +az\left( 5x^2+5y^2+5b^2-z^2\right)\left( \mathcal{R}_1+\mathcal{R}_2 \right) 
-\left( \mathcal{R}_1-\mathcal{R}_2 \right)\left[ a^2\left( x^2+y^2+b^2 \right) \right. \right. \notag \\
\left. \left. +\left( x^2+y^2+b^2+z^2\right)
\left(  2x^2+2y^2+2b^2-z^2 \right)\right] \right\} \mbox{,} \label{eq_phi_b2} \\
\rho_{(b)}=\rho_0+\varepsilon\rho_1 \mbox{,} \label{eq_rho_b}
\end{gather} 
where
\begin{gather}
\rho_1=\frac{\lambda_0yb^2}{4\pi a^2 \left( x^2+y^2+b^2\right)^3\mathcal{R}_1^5\mathcal{R}_2^5}
\left\{ \mathcal{R}_2^5 \left[ 8z^3\left( z+a \right)^5 \right. \right. \notag \\
\left. \left. +2z\left( z+a \right)^3 \left( x^2+y^2+b^2\right)\left( 13z^2+6az+3a^2 \right) \right. \right. \notag \\
\left. \left. +15z\left( z+a \right)\left( x^2+y^2+b^2\right)^2\left( 2z^2+2az+a^2 \right) \right. \right. \notag \\
\left. \left. +\left( x^2+y^2+b^2\right)^3\left( 14z^2+10az+5a^2 \right)+ 
2\left( x^2+y^2+b^2\right)^4 \right] \right. \notag \\
\left. -\mathcal{R}_1^5 \left[ 8z^3\left( z-a \right)^5+2z\left( z-a \right)^3
\left( x^2+y^2+b^2\right)\left( 13z^2-6az+3a^2 \right) \right. \right. \notag \\
\left. \left. +15z\left( z-a \right)\left( x^2+y^2+b^2\right)^2
\left( 2z^2-2az+a^2 \right) \right. \right. \notag \\
\left. \left. +\left( x^2+y^2+b^2\right)^3\left( 14z^2-10az+5a^2 \right)
+2\left( x^2+y^2+b^2\right)^4 \right] \right\} \mbox{,}
\end{gather}
and $\rho_0$ is given by (\ref{eq_rho_0}). Some level curves in the three orthogonal planes 
of the mass density $\bar{\rho}_{(b)}=\rho_{(b)}/(\lambda_0/a^2)$, equation (\ref{eq_rho_b}), 
are shown in Figs \ref{fig4}(a)--(c) with parameters $b/a=0.25$ and $\varepsilon=0.15$. 
The density contours near the origin are bent in the same manner as the thin line of 
Fig.\ \ref{fig1}(b). Two points of maximum are present near the $z$-axis and two points of minimum are located
antisymmetrically. The mass density may also become negative at these points with
increasing values of $\varepsilon$, which can rise up to $\varepsilon \approx 0.18$
for $b/a=0.25$. When the parameter $b/a$ is increased with fixed $\varepsilon$, the
point of minimum may disappear, the points of maximum approach the origin  and the 
isodensity contours become more circular, in the same manner as in the previous case.
Figs \ref{fig5}(a)--(c) display isopotential curves of the potential 
$\bar{\Phi}_{(b)}=\Phi_{(b)}/(G\lambda_0)$, equation (\ref{eq_phi_b2}), with parameters 
$b/a=0.25$ and $\varepsilon=0.15$. Here the minimum of potential is located at the origin. 
The components of the force $\mathbf{F}_{(b)}=-\nabla \Phi_{(b)}$
corresponding to the potential (\ref{eq_phi_b2}),  are also given in Appendix \ref{ap_A}. 

\section{Non-axisymmetric thin distributions of matter} \label{sec_nax}
Non-axisymmetric thin distributions of matter are of 
interest since many galaxies exhibit asymmetries in their discs (see e.g.\  
\cite{bls80,rs94} for observational evidences of asymmetries in disc galaxies).  
The potential of a thin bent bar may be used to generate non-axisymmetric thin 
distributions of matter. 
The procedure is similar to the one proposed by Kuzmin
\cite{k56} to find the gravity field of an axisymmetric disc. A source of gravitational 
field is placed below a plane $z=0$. Above the plane this gives a solution of the 
Laplace equation. The solution valid for $z \geq 0$ is then reflected with respect to
$z=0$ so as to give a symmetrical solution of the Laplace equation above and below 
the plane. This introduces a discontinuity in the normal derivative on $z=0$, which 
by the Poisson equation results in a surface density. Kuzmin applied his procedure on
a point mass. In general relativity, where the Schwarzschild solution in Weyl's
metric is represented by a finite bar with constant density, Kuzmin's method has been used
to construct `generalized Schwarzschild' discs \cite{b93a}. Some other examples of
general relativistic discs generated by Kuzmin's method can be found in
\cite{b93b,ll94,gl00,vl03}.

We will place the thin bent bar along the $x$-axis and then displace it at a distance $c>0$ 
below the plane $z=0$, and make the reflection with respect to that plane. This is 
equivalent to first make the replacements $z \rightarrow x$ and $x \rightarrow z$
in the potentials (\ref{eq_phi_a1})--(\ref{eq_phi_b1}), and further apply the 
transformation $z \rightarrow c+ \lvert z \rvert$. By Poisson equation, the surface density $\sigma$ on $z=0$
is given by \cite{bt08}
\begin{equation}
\sigma=\frac{1}{2\pi G} \left.  \frac{\partial \Phi}{\partial z} \right\rvert_{z=0} \mbox{.}
\end{equation}
For the transformed potentials (\ref{eq_phi_a1})--(\ref{eq_phi_b1}) we get, respectively, 
\begin{gather} 
\sigma_{(a)}=\frac{\lambda_0c\left[ \left( x+a \right)\mathrm{R}_2-\left( x-a \right)\mathrm{R}_1 \right]}
{2\pi \left( y^2+c^2 \right) \mathrm{R}_1\mathrm{R}_2}-\varepsilon\frac{\lambda_0cy}{2\pi a \left( y^2+c^2 \right)^2 
\mathrm{R}_1^3\mathrm{R}_2^3} \notag \\
\times \left\{ \mathrm{R}_1^3\left[ 2x\left( x^2+y^2+c^2 \right)^2 
-a\left( y^2+c^2 \right) \left( a^2-3ax+6x^2 \right) \right. \right. \notag \\
 \left. \left. -2ax^2\left( a^2-3ax+3x^2 \right) \right] -\mathrm{R}_2^3\left[ 2x\left( x^2+y^2+c^2 \right)^2 \right. \right. \notag \\
\left. \left. +a\left( y^2+c^2 \right) \left( a^2+3ax+6x^2 \right) +2ax^2\left( a^2+3ax+3x^2 \right) \right] \right\} \mbox{,} \label{eq_sigma_a} \\
\sigma_{(b)}=\frac{\lambda_0c\left[ \left( x+a \right)\mathrm{R}_2-\left( x-a \right)\mathrm{R}_1 \right]}
{2\pi \left( y^2+c^2 \right) \mathrm{R}_1\mathrm{R}_2}-\varepsilon \frac{\lambda_0cy}{2\pi a^2 \left( y^2+c^2 \right)^2 
\mathrm{R}_1^3\mathrm{R}_2^3} \notag \\
\times \left\{ \mathrm{R}_1^3\left[  2\left( x^2+y^2+c^2 \right)^3+3a \left( y^2+c^2 \right)^2 \left(
a-2x \right) \right. \right. \notag \\
\left. \left. -3ax\left( y^2+c^2 \right) \left( a^2-3ax+4x^2 \right)-2ax^3\left( a^2-3ax+3x^2 \right) \right] \right. \notag \\
\left. - \mathrm{R}_2^3\left[  2\left( x^2+y^2+c^2 \right)^3+3a \left( y^2+c^2 \right)^2 \left( 
a+2x \right) \right. \right. \notag \\
\left. \left. +3ax\left( y^2+c^2 \right) \left( a^2+3ax+4x^2 \right) +2ax^3\left( a^2+3ax+3x^2 \right) \right] \right\}
\mbox{,} \label{eq_sigma_b}
\end{gather}
where $\mathrm{R}_1=\sqrt{y^2+c^2+\left(x+a \right)^2}$ and $\mathrm{R}_2=\sqrt{y^2+c^2+\left(x-a \right)^2}$.  
Note that the first term in (\ref{eq_sigma_a}) and (\ref{eq_sigma_b}) is the surface density of a flat bar 
given by equation (11) of \cite{lm92}. Although they obtained the result by convolution of a flat Miyamoto-Nagai 
disc (which is in fact a Kuzmin disc) with a needle density, our procedure gives the same result.  

\begin{figure}
\centering
\includegraphics[scale=0.75]{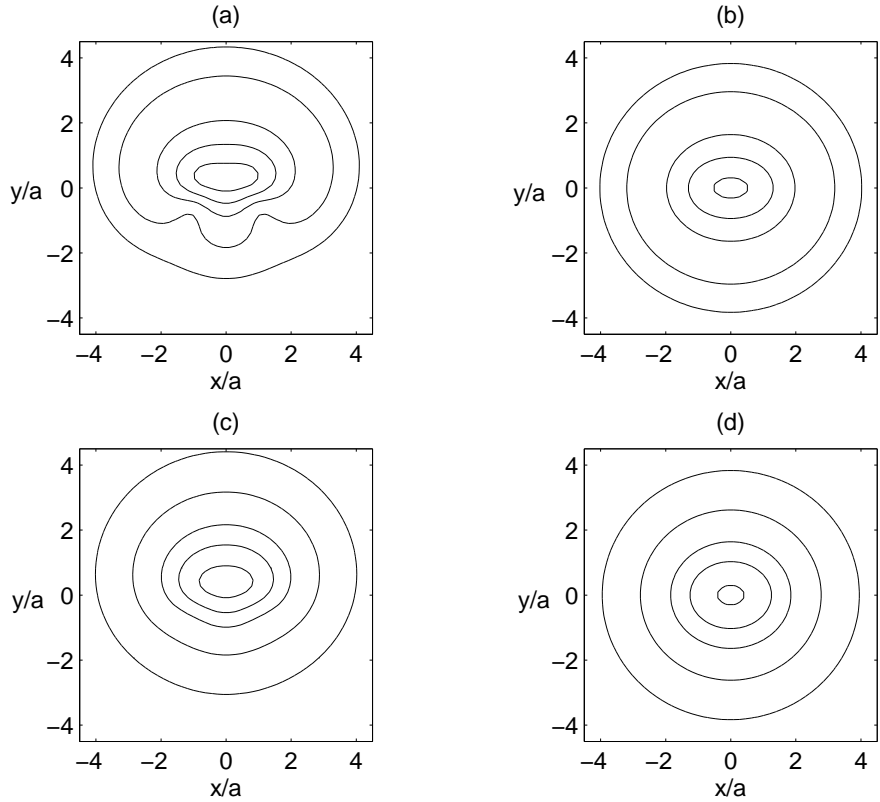}
\caption{(a)--(b) Isodensity contours of the surface density
$\bar{\sigma}_{(a)}=\sigma_{(a)}/(\lambda_0/a)$,
equation (\ref{eq_sigma_a}). (a) Parameters: $c/a=1$ and $\varepsilon=2$.
(b) Parameters: $c/a=1$ and $\varepsilon=0$. (c)--(d) Isopotential 
contours of the transformed potential $\bar{\Phi}_{(a)}=\Phi_{(a)}/(G\lambda_0)$ 
for the same parameters as in (a)--(b), respectively.} \label{fig6}
\end{figure}
\begin{figure}
\centering
\includegraphics[scale=0.65]{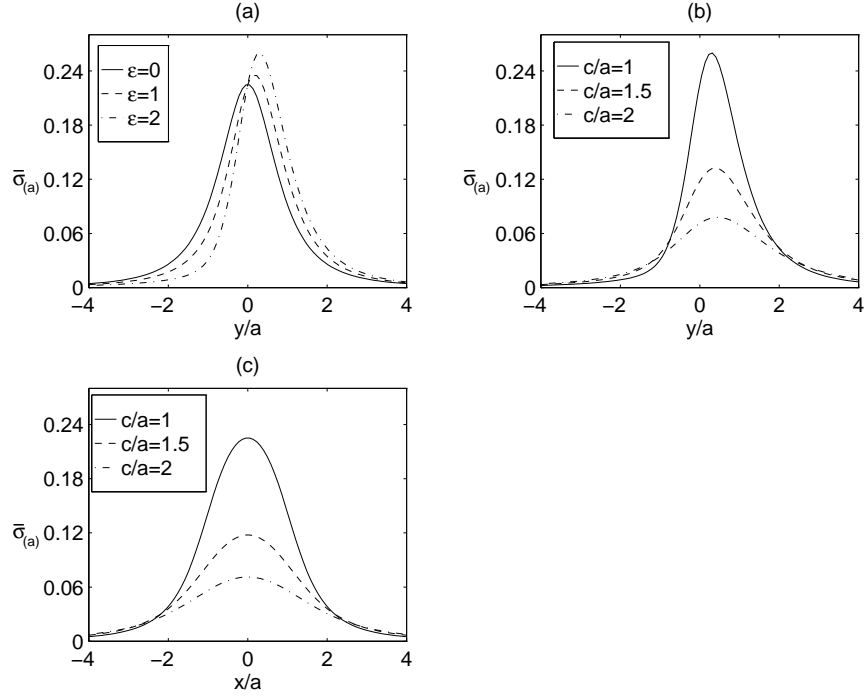}
\caption{(a)--(b) Profiles of the surface density  
$\bar{\sigma}_{(b)}=\sigma_{(b)}/(\lambda_0/a)$,
equation (\ref{eq_sigma_a}), along the $y$-axis, for different values of the parameters. 
(a) Parameters: $c/a=1$ and $\varepsilon=0$, $1$ and $2$. (b) Parameters:
$\varepsilon=2$ and $c/a=1$, $1.5$ and $2$. (c) Profiles of the same surface
density along the $x$-axis. Parameters: $c/a=1$, $1.5$ and $2$.} \label{fig7}
\end{figure}
\begin{figure}
\centering
\includegraphics[scale=0.75]{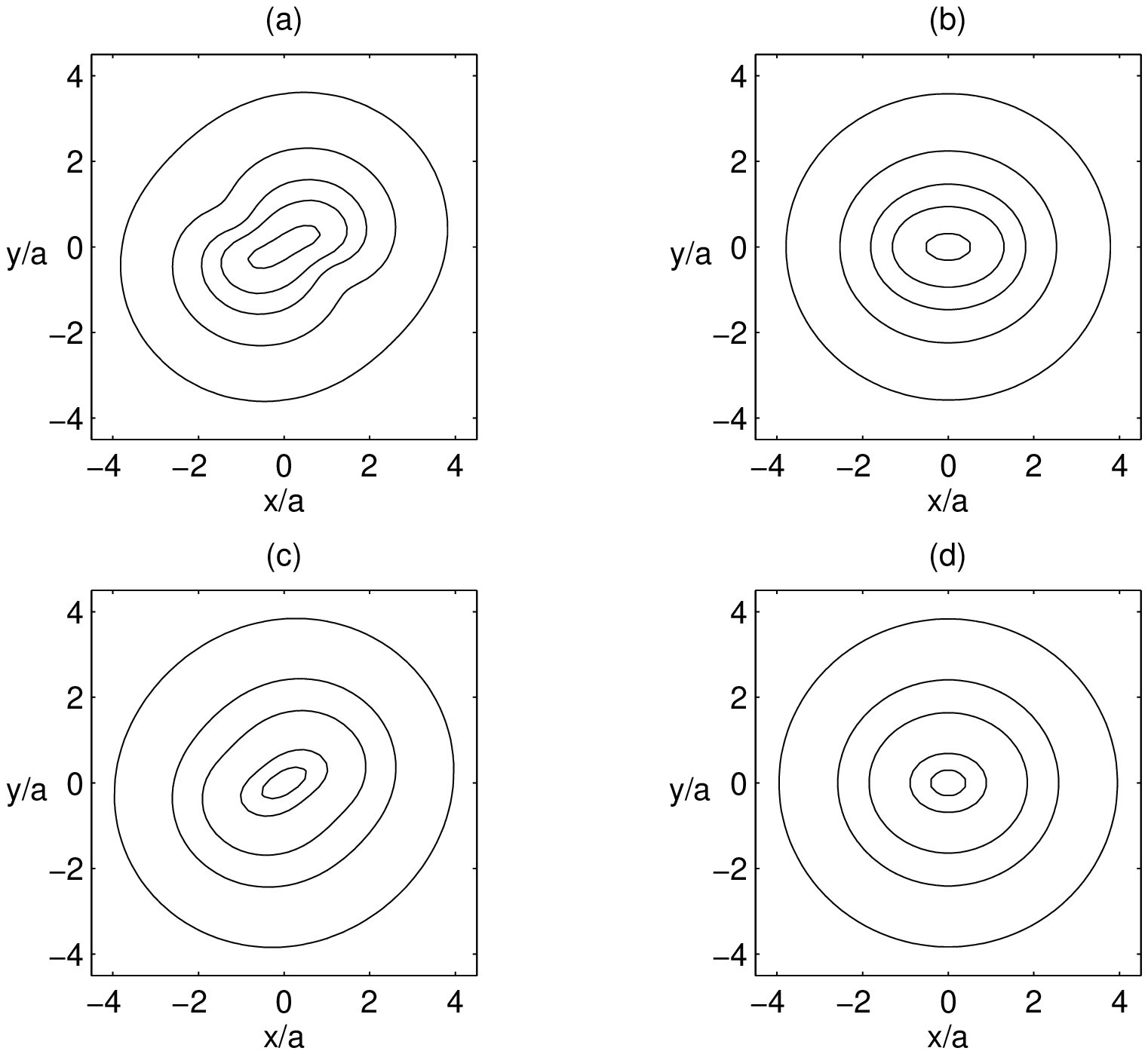}
\caption{(a)--(b) Isodensity contours of the surface density
$\bar{\sigma}_{(b)}=\sigma_{(b)}/(\lambda_0/a)$,
equation (\ref{eq_sigma_b}). (a) Parameters: $c/a=1$ and $\varepsilon=2$.
(b) Parameters: $c/a=1$ and $\varepsilon=0$. (c)--(d) Isopotential 
contours of the transformed potential $\bar{\Phi}_{(b)}=\Phi_{(b)}/(G\lambda_0)$
for the same parameters as in (a)--(b), respectively.} \label{fig8}
\end{figure}
Some contours of the dimensionless surface density $\bar{\sigma}_{(a)}=\sigma_{(a)}/(\lambda_0/a)$
are shown in Fig.\ \ref{fig6}(a), for $c/a=1$ and $\varepsilon=2$, and in
Fig.\ \ref{fig6}(b) for $c/a=1$ and $\varepsilon=0$. In both figures the 
level contours are plot for the same set of values. Figs \ref{fig6}(c) and (d)
display contours of the potential $\bar{\Phi}_{(a)}=\Phi_{(a)}/(G\lambda_0)$  for the
same parameters as in Figs \ref{fig6}(a) and (b), respectively. The isopotentials 
are also plot for the same set of values in both figures. Comparing the isodensity 
contours of the flat bent bar in Fig.\ \ref{fig6}(a) with the flat bar in Fig.\ \ref{fig6}(b), 
we note the asymetry with respect to the $x$-axis, with two points of minimum located symmetrically
with respect to the $y$-axis. For $c/a=1$, the surface density of the
flat bent bar is everywhere non-negative provided $\varepsilon \lessapprox 2.2$. The isopotential contours
of the flat bent bar are shifted along the $y$-axis, with a point of minimum at $x=0$ and $y=y_m$, which is the
zero of
\begin{multline}
\frac{2ay_m}{\left( y_m^2+c^2\right)\sqrt{y_m^2+c^2+a^2}}+\frac{\varepsilon}{a} \ln
\left( \frac{\sqrt{y_m^2+c^2+a^2}-a}{\sqrt{y_m^2+c^2+a^2}+a}\right) \\
+\frac{2\varepsilon \left[ \left( y_m^2+c^2 \right)^2+a^2\left( 2y_m^2+c^2\right)\right]}
{\left( y_m^2+c^2\right)\left( y_m^2+c^2+a^2\right)^{3/2}}=0 \mbox{,}
\end{multline}
an expression similar to (\ref{eq_y_m}). For the parameters $c/a=1$ and $\varepsilon=2$
we have $y_m/a \approx 0.376$.

In Figs \ref{fig7}(a)--(c) we display curves of the surface density
$\bar{\sigma}_{(a)}$ along the $y$- and $x$-axes  for some other values of the parameters $c/a$ and $\varepsilon$.
In Fig.\ \ref{fig7}(a) the parameter $c/a=1$ is kept constant. With increasing 
values of $\varepsilon$, the asymmetry of the density profile with respect to $y=0$ 
is enhanced. If we maintain $\varepsilon=2$ constant, as in Fig.\ \ref{fig7}(b),
the density distribution becomes somewhat less concentrated as the values of the parameter $c/a$ increase.
Fig.\ \ref{fig7}(c) shows curves of the surface density along the $x$-axis for the same values of  
$c/a$ as in Fig.\ \ref{fig7}(b). From (\ref{eq_sigma_a}), it is seen that for $y=0$ the profiles
of the surface density of the flat bent bar and of the flat undeformed bar are the same, since the
second term in (\ref{eq_sigma_a}) vanishes.

In Figs \ref{fig8}(a) and (b) we display some contours of the dimensionless surface density 
$\bar{\sigma}_{(b)}=\sigma_{(b)}/(\lambda_0/a)$, equation (\ref{eq_sigma_b}), 
for parameters $c/a=1$ and $\varepsilon=2$, and for $c/a=1$ and $\varepsilon=0$, respectively. 
In both figures the level contours are plot for the same set of values. Figs \ref{fig8}(c) and (d)
display contours of the potential $\bar{\Phi}_{(b)}=\Phi_{(b)}/(G\lambda_0)$, for the
same parameters as in Figs \ref{fig8}(a) and (b), respectively. The isopotentials 
are also plot for the same set of values in both figures. The surface density as well
as the potential for this model of flat bent bar are asymmetric with respect to the $x$-  
and $y$-axes, in contrast to the potential-density pair of the flat undeformed bar. 
For the parameter $c/a=1$, we find that the surface density of this model
flat bent bar is everywhere non-negative provided $\varepsilon \lessapprox 3.3$. 
The potential has a point of minimum at the origin. From (\ref{eq_sigma_b}) it is
straightforward to see that the surface density profiles along the $x$- and $y$-axes 
are the same for the flat bent and for the flat undeformed bars. 

 In Appendix \ref{ap_B}, we list the components of the force $\mathbf{F}_{(a)}=-\nabla \Phi_{(a)}$,
and of the force $\mathbf{F}_{(b)}=-\nabla \Phi_{(b)}$
on the plane $z=0$, after applying the transformations 
$z \rightarrow x$ and $x \rightarrow z$, then $z \rightarrow c+ \lvert z \rvert$,
on the potentials (\ref{eq_phi_a1})--(\ref{eq_phi_b1}).
\section{Discussion} \label{sec_dis}
We presented a method to bend a thin massive line when the curvature 
is small. The potential of the bent system is obtained from an expansion 
with respect to a small parameter. The procedure was then applied to a homogeneous 
bar with two examples of `bending functions'.  We showed that  if the `bending function'
can be expressed in terms of Legendre polynomials, then the potential of the bent bar is an
exact solution of the Laplace equation. Potential-density pairs for `softened'
bent bars were constructed by using a Plummer-like transformation. The resulting 
mass density distributions are non-negative everywhere for restricted values of the 
deformation parameter $\varepsilon$. We also used the potentials of the bent thin 
bars to construct planar distributions of matter without axial symmetry. Furthermore, 
non-negative surface density distributions that are non-symmetric 
with respect to one or both $x$- and $y$-axes can be found for restricted values of the 
deformation parameter $\varepsilon$.  

We would like to mention that the potential (\ref{eq_pert_l}) of a bent massive line contains only 
terms originated from the expansion with respect to $\varepsilon$ up to first order (dipole terms). 
In principle, higher order terms can be 
included, but we found that the explicit expressions, for instance  the quadrupole terms, are  
very cumbersome. In particular, the quadrupolar terms can be explicitly found using the 
same algorithm employed for the dipolar terms. 

A future work will be the study of orbits in the potentials of the `softened'
bent bars in an axisymmetric background with and without rotation and comparing the results with the 
works with undeformed bars, e.g. \cite{vf72,abmp83,pp83,pf84}.    
\section*{Acknowledgments}

DV thanks FAPESP for financial support, PSL thanks FAPESP and CNPq
for partial financial support. Authors also thank the anonymous referee for many comments 
and suggestions that improved this work. This research has 
made use of SAO/NASA's Astrophysics Data System Abstract Service, which is gratefully 
acknowledged.

\appendix
\section{Components of forces for bent bars} \label{ap_A}
The components of the force $\mathbf{F}_{(a)}$ corresponding to the potential (\ref{eq_phi_a2}), are
given by 
\begin{gather} 
F_{(a)x}=\frac{G\lambda_0x\left[ \left( z-a \right) \mathcal{R}_1 -\left( z+a \right) \mathcal{R}_2\right]}
{\left( x^2+y^2+b^2\right)\mathcal{R}_1\mathcal{R}_2}+\varepsilon \frac{G\lambda_0xy}
{a\left( x^2+y^2+b^2\right)^2\mathcal{R}_1^3\mathcal{R}_2^3} \notag \\
\times \left\{ \mathcal{R}_1^3 \left[ 2z \left( x^2+y^2+b^2+z^2\right)^2 
-a \left( x^2+y^2+b^2\right)\left( a^2-3az+6z^2 \right) \right. \right. \notag \\
\left. \left. -2az^2 \left( a^2-3az+3z^2 \right) \right] -
\mathcal{R}_2^3 \left[ 2z \left( x^2+y^2+b^2+z^2\right)^2 \right. \right. \notag \\
\left. \left.  +a \left( x^2+y^2+b^2\right)\left( a^2+3az+6z^2 \right)+2az^2 \left( a^2+3az+3z^2 \right) \right] \right\} \mbox{,} \\
F_{(a)y}= \frac{G\lambda_0y\left[ \left( z-a \right) \mathcal{R}_1 -\left( z+a \right) \mathcal{R}_2\right]}
{\left( x^2+y^2+b^2\right)\mathcal{R}_1\mathcal{R}_2} - \varepsilon \frac{G\lambda_0}{a} 
\ln \left( \frac{z-a+\mathcal{R}_2}{z+a+\mathcal{R}_1} \right) \notag \\
+ \varepsilon \frac{G\lambda_0}
{a\left( x^2+y^2+b^2\right)^2\mathcal{R}_1^3\mathcal{R}_2^3} \left\{ \mathcal{R}_1^3 \Bigl[ 
z \left( x^2+y^2+b^2+z^2\right)^2  \left( y^2-x^2-b^2 \right) \Bigr. \right. \notag \\
 \left. \Bigl. -a \left( x^2+y^2+b^2+z^2\right) \left[ 
3z^2 \left( y^2-x^2-b^2 \right)+ \left( x^2+y^2+b^2\right)^2 \right] \Bigr. \right. \notag \\
\left. \Bigl.-a^2 \left( x^2+y^2+b^2\right) \left[ a \left( x^2+2y^2+b^2\right) 
-z\left( x^2+4y^2+b^2 \right) \right] \Bigr. \right. \notag \\
\left. \Bigl. -a^2z^2 \left( y^2-x^2-b^2 \right) \left( a-3z \right) \Bigr] -  \mathcal{R}_2^3 \Bigl[ 
z \left( x^2+y^2+b^2+z^2\right)^2 \left( y^2-x^2-b^2 \right) \Bigr. \right. \notag \\
\left. \Bigl. +a \left( x^2+y^2+b^2+z^2\right) \left[ 
3z^2 \left( y^2-x^2-b^2 \right)+ \left( x^2+y^2+b^2\right)^2 \right] \Bigr. \right. \notag \\
\left. \Bigl. +a^2 \left( x^2+y^2+b^2\right) \left[ 
a \left( x^2+2y^2+b^2\right)+z\left( x^2+4y^2+b^2 \right) \right] \Bigr. \right. \notag \\ 
\left. \Bigl. +a^2z^2 \left( y^2-x^2-b^2 \right) \left( a+3z \right) \Bigr] \right\} \mbox{,} \\
F_{(a)z}=\frac{G\lambda_0\left( \mathcal{R}_2-\mathcal{R}_1\right)}{\mathcal{R}_1\mathcal{R}_2} + 
\varepsilon \frac{G\lambda_0y}{a\left( x^2+y^2+b^2\right)\mathcal{R}_1^3\mathcal{R}_2^3} \left\{ 
\mathcal{R}_2^3 \left[ 2 \left( x^2+y^2+b^2+z^2\right)^2 \right. \right. \notag \\ 
\left. \left. +3a \left( x^2+y^2+b^2\right) \left( a+2z \right) 
+2az \left( a^2+3az+3z^2 \right) \right] \right. \notag \\
\left. - \mathcal{R}_1^3 \left[ 2 \left( x^2+y^2+b^2+z^2\right)^2
+3a \left( x^2+y^2+b^2\right) \left( a-2z \right) \right. \right. \notag \\ 
\left. \left. - 2az \left( a^2-3az+3z^2 \right) \right] \right\} \mbox{.}
\end{gather}

The components of the force $\mathbf{F}_{(b)}$ corresponding to the potential (\ref{eq_phi_b2}), 
 are given by
\begin{gather}
F_{(b)x}=\frac{G\lambda_0x\left[ \left( z-a \right) \mathcal{R}_1 -\left( z+a \right) \mathcal{R}_2\right]}
{\left( x^2+y^2+b^2\right)\mathcal{R}_1\mathcal{R}_2}+\varepsilon \frac{G\lambda_0xy}
{a^2\left( x^2+y^2+b^2\right)^2\mathcal{R}_1^3\mathcal{R}_2^3} \notag \\
\left\{ \mathcal{R}_1^3 \left[ 
2 \left( x^2+y^2+b^2+z^2\right)^3+3a \left( x^2+y^2+b^2\right)^2 \left( a-2z \right) \right. \right. \notag \\
\left. \left. -3az \left( x^2+y^2+b^2\right)\left( a^2-3az+4z^2 \right) -2az^3 \left( a^2-3az+3z^2 \right) \right] \right. \notag \\
\left. -\mathcal{R}_2^3 \left[ 2 \left( x^2+y^2+b^2+z^2\right)^3+3a \left( x^2+y^2+b^2\right)^2 \left( a+2z \right) 
\right. \right. \notag \\
\left. \left. +3az \left( x^2+y^2+b^2\right)\left( a^2+3az+4z^2 \right) +2az^3 \left( a^2+3az+3z^2 \right) \right] 
\right\} \mbox{,} \\
F_{(b)y}= \frac{G\lambda_0y\left[ \left( z-a \right) \mathcal{R}_1 -\left( z+a \right) \mathcal{R}_2\right]}
{\left( x^2+y^2+b^2\right)\mathcal{R}_1\mathcal{R}_2} -3\varepsilon \frac{G\lambda_0z}{a^2}
\ln \left( \frac{z-a+\mathcal{R}_2}{z+a+\mathcal{R}_1} \right) \notag \\
+ \varepsilon \frac{G\lambda_0}
{a^2\left( x^2+y^2+b^2\right)^2\mathcal{R}_1^3\mathcal{R}_2^3} \left\{ \mathcal{R}_1^3 \Bigl[ 
\left( x^2+y^2+b^2+z^2\right)^2 \Bigr. \right. \notag \\
\left. \Bigl.  \times \left[ 2 \left( x^2+y^2+b^2\right)\left( x^2+2y^2+b^2\right)+z^2 
\left( y^2-x^2-b^2\right) \right] \Bigr. \right. \notag \\
\left. \Bigl. -az^3 \left( y^2-x^2-b^2\right) \left( a^2-3az+3z^2 \right)+ 
a^4 \left( x^2+y^2+b^2\right)^2  \Bigr. \right. \notag \\
\left. \Bigl. -3a\left( x^2+y^2+b^2\right)^2 \left[ z\left( 3x^2+5y^2+3b^2\right) 
-a\left( x^2+2y^2+b^2\right)\right]  \Bigr. \right. \notag \\
\left. \Bigl. -az\left( x^2+y^2+b^2\right) \left[ -3az \left( 4x^2+7y^2+4b^2\right)
+6z^2\left( x^2+3y^2+b^2\right) \right. \Bigr. \right. \notag \\ 
\left. \Bigl. \left. +a^2\left( 7x^2+10y^2+7b^2\right)\right] \Bigr]-\mathcal{R}_2^3 \Bigl[
\left( x^2+y^2+b^2+z^2\right)^2 \Bigr. \right. \notag \\
\left. \Bigl. \times \left[ 2 \left( x^2+y^2+b^2\right)\left( x^2+2y^2+b^2\right) 
+z^2\left( y^2-x^2-b^2\right) \right] \Bigr. \right. \notag \\
\left. \Bigl. +az^3 \left( y^2-x^2-b^2\right) \left( a^2+3az+3z^2 \right)
+a^4 \left( x^2+y^2+b^2\right)^2 \Bigr. \right. \notag \\
\left. \Bigl.  +3a\left( x^2+y^2+b^2\right)^2 \left[ z\left( 3x^2+5y^2+3b^2\right)  
+a\left( x^2+2y^2+b^2\right)\right] \Bigr. \right. \notag \\
\left. \Bigl. +az\left( x^2+y^2+b^2\right) \left[ 3az \left( 4x^2+7y^2+4b^2\right)
+6z^2\left( x^2+3y^2+b^2\right) \right. \Bigr. \right. \notag \\ 
\left. \Bigl. \left. +a^2\left( 7x^2+10y^2+7b^2\right)\right] \Bigr] \right\} \mbox{,} \\
F_{(b)z}=\frac{G\lambda_0\left( \mathcal{R}_2-\mathcal{R}_1\right)}{\mathcal{R}_1\mathcal{R}_2}
-3\varepsilon \frac{G\lambda_0y}{a^2} \ln \left( \frac{z-a+\mathcal{R}_2}{z+a+\mathcal{R}_1} \right) \notag \\
-\varepsilon \frac{G\lambda_0y}{a^2\left( x^2+y^2+b^2\right)\mathcal{R}_1^3\mathcal{R}_2^3}
\left\{ \mathcal{R}_1^3 \left[ 3z \left( x^2+y^2+b^2+z^2\right)^2 \right. \right. \notag \\
\left. \left. -a\left( x^2+y^2+b^2\right) 
\left( 6z^2+3az-4a^2 \right)+3a\left( x^2+y^2+b^2\right)^2 \right. \right. \notag \\
\left. \left.  -3az^2 \left( 3z^2-3az+a^2 \right) \right] 
- \mathcal{R}_2^3 \left[ 3z \left( x^2+y^2+b^2+z^2\right)^2 \right. \right. \notag \\
\left. \left. +a\left( x^2+y^2+b^2\right)
\left( 6z^2-3az-4a^2 \right)-3a\left( x^2+y^2+b^2\right)^2 \right. \right. \notag \\
\left. \left. +3az^2 \left( 3z^2+3az+a^2 \right)  \right] \right\} \mbox{.}
\end{gather}
In all equations, we have $\mathcal{R}_1=\sqrt{x^2+y^2+b^2+\left(z+a \right)^2}$, and \newline
$\mathcal{R}_2=\sqrt{x^2+y^2+b^2+\left(z-a \right)^2}$. 

\section{Components of forces for non-axisymmetric thin distributions of matter} \label{ap_B} 
The components of the force $\mathbf{F}_{(a)}=-\nabla \Phi_{(a)}$,
on the plane $z=0$, after applying the transformations 
$z \rightarrow x$ and $x \rightarrow z$, then $z \rightarrow c+ \lvert z \rvert$,
on the potential (\ref{eq_phi_a1}), read
\begin{gather}
\mathrm{F}_{(a)x}=\frac{G\lambda_0\left( \mathrm{R}_2-\mathrm{R}_1\right)}{\mathrm{R}_1\mathrm{R}_2} + 
\varepsilon \frac{G\lambda_0y}{a\left( y^2+c^2\right)\mathrm{R}_1^3\mathrm{R}_2^3} \left\{ 
\mathrm{R}_2^3 \left[ 2 \left( x^2+y^2+c^2\right)^2 \right. \right. \notag \\
\left. \left. +3a \left( y^2+c^2\right) \left( a+2x \right) +2ax \left( a^2+3ax+3x^2 \right) \right] 
 - \mathrm{R}_1^3 \left[ 2 \left( x^2+y^2+c^2\right)^2 \right. \right. \notag \\
\left. \left. +3a \left( y^2+c^2\right) \left( a-2x \right)- 2ax \left( a^2-3ax+3x^2 \right) \right] \right\} \mbox{,} \\
\mathrm{F}_{(a)y}= \frac{G\lambda_0y\left[ \left( x-a \right) \mathrm{R}_1 -\left( x+a \right) \mathrm{R}_2\right]}
{\left( y^2+c^2\right)\mathrm{R}_1\mathrm{R}_2} - \varepsilon \frac{G\lambda_0}{a} 
\ln \left( \frac{x-a+\mathrm{R}_2}{x+a+\mathrm{R}_1} \right) \notag \\
+ \varepsilon \frac{G\lambda_0}
{a\left( y^2+c^2\right)^2\mathrm{R}_1^3\mathrm{R}_2^3} \left\{ \mathrm{R}_1^3 \Bigl[ 
x \left( x^2+y^2+c^2\right)^2\left( y^2-c^2 \right) \Bigr. \right. \notag \\
\left. \Bigl. -a \left( x^2+y^2+c^2\right) \left[
3x^2 \left( y^2-c^2 \right)+ \left( y^2+c^2\right)^2 \right] \Bigr. \right. \notag \\
\left. \Bigl. -a^2 \left( y^2+c^2\right) \left[ 
a \left( 2y^2+c^2\right)-x\left( 4y^2+c^2 \right) \right] -a^2x^2 \left( y^2-c^2 \right) \left( a-3x \right)
\Bigr] \right. \notag \\
\left. - \mathrm{R}_2^3 \Bigl[ x \left( x^2+y^2+c^2\right)^2 \left( y^2-c^2 \right) +a \left( x^2+y^2+c^2\right) \left[
3x^2 \left( y^2-c^2 \right)+ \left( y^2+c^2\right)^2 \right] \Bigr. \right. \notag \\
\left. \Bigl. +a^2 \left( y^2+c^2\right) \left[ a \left( 2y^2+c^2\right)+x\left( 4y^2+c^2 \right) \right] 
 +a^2x^2 \left( y^2-c^2 \right) \left( a+3x \right) \Bigr] \right\} \mbox{.}
\end{gather}

The components of the force $\mathbf{F}_{(b)}=-\nabla \Phi_{(b)}$,
on the plane $z=0$, after applying the transformations 
$z \rightarrow x$ and $x \rightarrow z$, then $z \rightarrow c+ \lvert z \rvert$,
on the potential (\ref{eq_phi_b1}), read
\begin{gather} 
\mathrm{F}_{(b)x}=\frac{G\lambda_0\left( \mathrm{R}_2-\mathrm{R}_1\right)}{\mathrm{R}_1\mathrm{R}_2}
-3\varepsilon \frac{G\lambda_0y}{a^2} \ln \left( \frac{x-a+\mathrm{R}_2}{x+a+\mathrm{R}_1} \right)- 
\varepsilon \frac{G\lambda_0y}{a^2\left( y^2+c^2\right)\mathrm{R}_1^3\mathrm{R}_2^3} \notag \\
\times \left\{ \mathrm{R}_1^3 \left[ 3x \left( x^2+y^2+c^2\right)^2 -a\left( y^2+c^2\right) 
\left( 6x^2+3ax-4a^2 \right)+3a\left( y^2+c^2\right)^2 \right. \right. \notag \\
\left. \left. -3ax^2 \left( 3x^2-3ax+a^2 \right) \right]
- \mathrm{R}_2^3 \left[ 3x \left( x^2+y^2+c^2\right)^2 \right. \right. \notag \\
\left. \left. +a\left( y^2+c^2\right)
\left( 6x^2-3ax-4a^2 \right)-3a\left( y^2+c^2\right)^2+3ax^2 \left( 3x^2+3ax+a^2 \right)
\right] \right\} \mbox{,} \\
\mathrm{F}_{(b)y}= \frac{G\lambda_0y\left[ \left( x-a \right) \mathrm{R}_1 -\left( x+a \right) \mathrm{R}_2\right]}
{\left( y^2+c^2\right)\mathrm{R}_1\mathrm{R}_2} -3\varepsilon \frac{G\lambda_0x}{a^2}
\ln \left( \frac{x-a+\mathrm{R}_2}{x+a+\mathrm{R}_1} \right) \notag \\
+ \varepsilon \frac{G\lambda_0}
{a^2\left( y^2+c^2\right)^2\mathrm{R}_1^3\mathrm{R}_2^3} \left\{ \mathrm{R}_1^3 \Bigl[ 
\left( x^2+y^2+c^2\right)^2  \left[ 2 \left( y^2+c^2\right)\left( 2y^2+c^2\right) \right. \Bigr. \right. \notag \\
\left. \Bigl. \left. +x^2 \left( y^2-c^2\right) \right] -ax^3 \left( y^2-c^2\right) \left( a^2-3ax+3x^2 \right)+ 
a^4 \left( y^2+c^2\right)^2 \right. \Bigr. \notag \\
\left. \Bigl.  -3a\left( y^2+c^2\right)^2 \left[ x\left( 5y^2+3c^2\right) -a\left( 2y^2+c^2\right)\right] \right. \Bigr. \notag \\
\left. \Bigl. -ax\left( y^2+c^2\right) \left[ -3ax \left( 7y^2+4c^2\right)
+6x^2\left( 3y^2+c^2\right)+a^2\left( 10y^2+7c^2\right)\right] \Bigr] \right. \notag \\
\left. -\mathrm{R}_2^3 \Bigl[ \left( x^2+y^2+c^2\right)^2 \left[ 2 \left( y^2+c^2\right)\left( 2y^2+c^2\right)
+x^2\left( y^2-c^2\right) \right] \Bigr. \right. \notag \\
\left. \Bigl. +ax^3 \left( y^2-c^2\right) \left( a^2+3ax+3x^2 \right)+a^4 \left( y^2+c^2\right)^2  \Bigr. \right. \notag \\
\left. \Bigl. +3a\left( y^2+c^2\right)^2 \left[ x\left( 5y^2+3c^2\right) 
+a\left( 2y^2+c^2\right)\right] \Bigr. \right. \notag \\
\left. \Bigl. +ax\left( y^2+c^2\right) \left[ 3ax \left( 7y^2+4c^2\right)
+6x^2\left( 3y^2+c^2\right)+a^2\left( 10y^2+7c^2\right)\right] \Bigr] \right\} \mbox{.} 
\end{gather} 
In all equations, we have $\mathrm{R}_1=\sqrt{y^2+c^2+\left(x+a \right)^2}$ and \newline 
$\mathrm{R}_2=\sqrt{y^2+c^2+\left(x-a \right)^2}$. 

\begin{thebibliography}{99}
\bibitem{sw93} Sellwood J. A., Wilkinson A., 1993, Rep. Progress Phys., 56, 173 
\bibitem{bm98} Binney J., Merrifield M., 1998, Galactic Astronomy, Princeton Univ. Press, Princeton, NJ
\bibitem{dev63} de Vaucouleurs G., 1963, ApJS, 8, 31
\bibitem{kn99} Knapen J. H., 1999, in Beckman J. E., Mahoney T. J., eds, ASP Conf. Ser. Vol. 187, 
The Evolution of Galaxies on Cosmological Timescales. Astron. Soc. Pac., San Francisco, p. 72
\bibitem{esk00} Eskridge P. B. et al., 2000, AJ, 119, 536
\bibitem{ksp00} Knapen J. H., Shlosman I., Peletier R. F., 2000, ApJ, 529, 93
\bibitem{fr66} Freeman K. C., 1966, MNRAS, 134, 15
\bibitem{d65} Danby J. M. A., 1965, AJ, 70, 501
\bibitem{m75} Michalodimitrakis M., 1975, Ap\&SS, 33, 421
\bibitem{vf72} de Vaucouleurs G., Freeman K. C., 1972, Vistas Astron., 14, 163
\bibitem{abmp83} Athanassoula E., Bienayme O., Martinet L., Pfenniger D., 1983, A\&A, 127, 349
\bibitem{pp83} Papayannopoulos T., Petrou M., 1983, A\&A, 119, 21
\bibitem{pf84} Pfenniger D., 1984, A\&A, 134, 373
\bibitem{fe77} Ferrers N. M., 1877, Q. J. Pure Applied Math., 14, 1
\bibitem{lm92} Long K., Murali C., 1992, ApJ, 397, 44
\bibitem{llak99} Lee C. W., Lee H. M., Ann H. B., Kwon K. H., 1999, ApJ, 513, 242 
\bibitem{al00} Ann H. B., Lee H. M., 2000, J. Korean Astron. Soc., 33, 1
\bibitem{al04} Ann H. B., Lee H. M., 2004, ApJ, 613, L105
\bibitem{at05} Ann H. B., Thakur P., 2005, ApJ, 620, 197
\bibitem{taj09} Thakur P., Ann H. B., Jiang I.-G., 2009, ApJ, 693, 586
\bibitem{bt08} Binney J., Tremaine S., 2008,
Galactic Dynamics, 2nd edn. Princeton Univ. Press, Princeton, NJ
\bibitem{ss88} Sellwood J. A., Sparke L. S., 1988, MNRAS, 231, 25p
\bibitem{ee89} Elmegreen B. G., Elmegreen D. M., 1989, ApJ, 342, 677
\bibitem{bl01} Block D. L., Puerari I., Knapen J. H., 
Elmegreen B. G., Buta R., Stedman S., Elmegreen D. M., 2001, A\&A, 375, 761
\bibitem{bl04} Block D. L., Buta R.,  Knapen J. H.,  
Elmegreen D. M., Elmegreen B. G.,  Puerari I., 2004, AJ, 128, 183
\bibitem{bu09} Buta R.,  Knapen J. H., Elmegreen B. G.,   
Salo H., Laurikainen E., Elmegreen D. M., Puerari I., Block D. L., 2009, AJ, 137, 4487
\bibitem{slbk10} Salo H., Laurikainen E., 
Buta R., Knapen J. H., 2010, ApJ, 715, L56
\bibitem{ath92} Athanassoula E., 1992, MNRAS, 259, 345
\bibitem{w94} Wada K., 1994, PASJ, 46, 165
\bibitem{eg97} Englmaier P., Gerhard O. E., 1997, MNRAS, 287, 57
\bibitem{fwh98} Fukuda H., Wada K., Habe A., 1998, MNRAS, 295, 463
\bibitem{es00} Englmaier P., Shlosman I., 2000, ApJ, 528, 677
\bibitem{pa00} Patsis P. A., Athanassoula E., 2000, A\&A, 358, 45
\bibitem{ma02}  Maciejewski W., Teuben P. J., 
Sparke L. S., Stone J. M., 2002, MNRAS, 329, 502
\bibitem{ma03} Maciejewski W., 2003, in Boily C. M., Patsis P., Portegies Zwart S., Spurzem R., 
Theis C., eds, Proc. JENAM 2002, Galactic and Stellar Dynamics.  EDP Sciences, Les Ulis, p. 3 
\bibitem{k56} Kuzmin G. G., 1956, AZh, 33, 27
\bibitem{l99} Letelier P. S., 1999, Classical Quantum Gravity, 16, 1207
\bibitem{bls80} Baldwin J. E., Lynden Bell D., Sancisi R., 1980, MNRAS, 193, 313
\bibitem{rs94} Richter O. -G., Sancisi R., 1994, A\&A, 290, L9
\bibitem{b93a} Bi\v{c}\'{a}k J., Lynden Bell D., Katz J., 1993, Phys. Rev. D, 47, 4334
\bibitem{b93b} Bi\v{c}\'{a}k J., Lynden Bell D., Pichon C., 1993, MNRAS, 265, 126
\bibitem{ll94} Lemos J. P. S., Letelier P. S., 1994, Phys. Rev. D, 49, 5135
\bibitem{gl00} Gonz\'alez G. A., Letelier P. S., 2000, Phys. Rev. D, 62, 064025
\bibitem{vl03} Vogt D., Letelier P. S., 2003, Phys. Rev. D, 68, 084010
\end{thebibliography}
\end{document}